\documentclass[aip,reprint]{revtex4-1}
\usepackage{graphicx}
\usepackage{amsmath, cases}
\usepackage{dcolumn}
\usepackage{bm}
\usepackage{ulem}
\usepackage{times}
\usepackage{amssymb}
\usepackage{epsfig}
\usepackage{xcolor}
\usepackage[T1]{fontenc}
\usepackage{mathtools, cases}   

\draft 

\begin{document}


\title{Generalized mode-coupling theory of the glass transition. I. Numerical results for Percus-Yevick hard spheres} 



\author{Chengjie Luo}
\email[Electronic mail: ]{C.Luo@tue.nl}
\affiliation{Theory of Polymers and Soft Matter, Department of Applied Physics,
	Eindhoven University of Technology, P.O. Box 513, 5600MB Eindhoven, The Netherlands}
\author{Liesbeth M.~C.~Janssen}
\email[Electronic mail: ]{L.M.C.Janssen@tue.nl}
\affiliation{Theory of Polymers and Soft Matter, Department of Applied Physics,
	Eindhoven University of Technology, P.O. Box 513, 5600MB Eindhoven, The Netherlands}


\date{\today}

\begin{abstract}
Mode-coupling theory (MCT) constitutes one of the few first-principles-based approaches to
describe the physics of the glass transition, but the theory's inherent approximations 
compromise its accuracy in the activated glassy regime. 
Here we show that microscopic generalized mode-coupling theory (GMCT),
a recently proposed hierarchical framework to systematically improve upon standard MCT, provides
a promising pathway toward a more accurate first-principles description of glassy dynamics. 
We present a comprehensive numerical analysis for Percus-Yevick hard spheres by performing 
explicitly wavenumber- and time-dependent GMCT calculations up to sixth order. 
Specifically, we calculate the location of the critical point, the associated non-ergodicity parameters, 
the time-dependent dynamics of the density correlators at both absolute and reduced packing fractions,
and we test several universal scaling relations in the $\alpha$- and $\beta$-relaxation regimes.
It is found that higher-order GMCT can successfully remedy some of standard MCT's pathologies, 
including an underestimation of the critical glass transition density and an overestimation of 
the hard-sphere fragility. Furthermore, we numerically demonstrate that the celebrated scaling laws of
standard MCT are preserved in GMCT at all closure levels, and that the predicted critical exponents 
manifestly improve as more levels are incorporated in the GMCT hierarchy. 
Although formally the GMCT equations should be solved up to infinite order to reach full convergence, our finite-order GMCT calculations 
unambiguously reveal a uniform convergence pattern for the dynamics. We thus argue that GMCT can provide a feasible and controlled means 
to bypass MCT's main uncontrolled approximation, offering hope for the future development of a quantitative first-principles theory of 
the glass transition.
\end{abstract}

\pacs{}

\maketitle 


\section{Introduction}

Understanding the physics of the glass transition is still one of the grand challenges
in condensed matter science.\cite{debenedetti2001supercooled,berthier2011theoretical} 
One of the most puzzling features of vitrification is that, upon supercooling
or compression, the relaxation dynamics of a glass-forming material slows down
by many orders of magnitude, while the microstructure undergoes only minute
changes. Moreover, not all materials vitrify in the same manner: so-called
strong glass-formers solidify rather gradually following an Arrhenius law,
whereas fragile materials exhibit a super-Arrhenius growth of the relaxation
time.  These differences in fragility imply that any universal theory of the
glass transition must be able to account for non-trivial material-dependent
properties.  Many theories and theoretical models have been proposed in the
past decades to rationalize this glassy phenomenology,\cite{berthier2011theoretical,kirkpatrick2015colloquium,ritort2003glassy,
adam1965temperature, royall2015role, ediger2000spatially,tarjus2011overview,biroli2013perspective}
but no theory to date can accurately predict the glass transition point, i.e.\
the temperature or density at which a supercooled liquid enters the non-ergodic glassy
state, the fragility, and the fully time-dependent relaxation dynamics on the
sole basis of a material's microstructure. More generally, a quantitative
first-principles framework to account for all relevant features of glass
formation is still lacking. 

The mode-coupling theory of the glass transition (MCT)\cite{gotze2008complex,leutheusser1984dynamical,bengtzelius1984dynamics,
reichman2005mode,janssen2018mode}
is essentially the only theory founded on purely first principles that can partially explain the
complex dynamics of glass-forming liquids.  Briefly, MCT starts from
the \textit{exact} equation of motion for the two-point density correlation
function $F(k,t)$--a microscopic probe for the structural relaxation dynamics
at a certain wavenumber $k$ and time $t$. This equation is governed by a memory
function that contains, to leading order, four-point dynamic density correlations; MCT
subsequently approximates these multi-point correlators as a product of
$F(k,t)$'s, resulting in a self-consistent equation that only requires the
material-dependent static structure factor $S(k)$ as input. Despite MCT's uncontrolled
factorization approximation of the memory function, however, the theory has
been remarkably successful in describing several non-trivial features of glass
formation.  These include the prediction of the two-step decay of $F(k,t)$ via
so-called $\beta$- and $\alpha$-relaxation processes, respectively, a physically
intuitive picture of the dramatic dynamical slowdown in terms of the cage
effect, non-trivial scaling laws in the $\beta$-relaxation regime, stretched
exponential decay and a time-temperature superposition principle in the
$\alpha$-relaxation regime, as well as complex glassy reentrant phenomena.\cite{pham2002multiple, berthier2010increasing}
Furthermore, the non-ergodicity parameters at the glass transition are generally
consistent with experimental data.\cite{van1991nonergodicity} 

However, MCT also suffers from several pathologies. Notably, MCT predicts a
spurious glass transition that is typically much higher (lower) than the
experimental glass transition temperature $T_g$ (packing fraction
$\varphi_g$).\cite{reichman2005mode} In general, the theory is therefore only
quantitatively accurate in the mildly supercooled regime, and efforts to extend
the quantitative applicability of MCT often rely on (\textit{ad hoc}) rescaling
procedures.  Moreover, in its standard form, MCT cannot accurately account for
the emergence of dynamical heterogeneity\cite{biroli2006inhomogeneous} and violation of the
Stokes-Einstein relation in strongly supercooled liquids.  The concept of
fragility is also not adequately captured by the theory.
In fact, MCT strictly predicts a power-law
divergence of the relaxation time, which may account for fragile behavior,
but is inconsistent with e.g.\ the empirical Vogel-Fulcher-Tamman (VFT) law\cite{berthier2011theoretical} and 
the Arrhenius behavior of strong materials.  Finally,
although MCT is often interpreted as a mean-field framework, the theory does
not become exact in the mean-field limit of infinite spatial
dimensions.\cite{ikeda2010mode,schmid2010glass,maimbourg2016solution} These are the reasons that from
more panoramic and thermodynamical viewpoints, such as the Random First Order
Transition Theory (RFOT),\cite{wolynes2012structural,berthier2011theoretical} the MCT transition only describes a crossover or
avoided transition in the dynamics. It must be noted, however, that the crossover between MCT and the activated regime is not clear and that some recent studies suggest that 
thermal activation is already at play in the temperature regime usually described by MCT.\cite{berthier2011theoretical}

As MCT is based entirely on first principles, the above problems of the theory
can all be traced back to MCT's uncontrolled, though well-defined,
approximations--in particular the factorization of four-point dynamical density
correlations.  A promising approach to improve MCT in a \textit{controlled}
manner is the so-called generalized mode-coupling theory (GMCT), a framework that was
first proposed by Szamel in 2003. Briefly, GMCT seeks to develop a new and
formally exact equation of motion for the unknown memory function in $F(k,t)$;
this new equation is governed by six-point correlations, which in turn are
dominated by eight-point correlations, and so on. This culminates into a
hierarchy of coupled integro-differential equations which may be closed (self-consistently) at
arbitrary order. Thus, GMCT allows one to postpone the factorization
approximation to a higher
level.\cite{szamel2003colloidal,wu2005high,janssen2015microscopic}

Szamel\cite{szamel2003colloidal} and Wu and Cao\cite{wu2005high} showed that
the predicted critical point $\varphi^c$ for glassy hard spheres indeed
systematically improves by including one or two additional levels in the GMCT
hierarchy, respectively, using only the static structure factor as input.  A
more recent GMCT study, which included up to three additional
levels,\cite{janssen2015microscopic} demonstrated that the theory's predictions
for the fully time-dependent microscopic dynamics of weakly polydisperse hard spheres also
converge to the empirical data, at least for the first few decades of
structural relaxation.  Overall, these calculations indicate that increasing
the closure level of the GMCT hierarchy generally leads to more liquid-like
(i.e.\ faster relaxation) dynamics as compared to standard MCT predictions \textit{at
the same density}, implying that higher-order correlations introduce more
ergodicity-restoring relaxation channels. This finding is encouraging,
considering that standard MCT generally overestimates the glassiness of a
material.  Finally, several wavevector-independent schematic GMCT models up to
infinite order revealed that GMCT should be mathematically capable of
accounting for different degrees of fragility, ranging from strong to fragile,\cite{janssen2014relaxation}
as well as a strictly avoided dynamical glass
transition.\cite{mayer2006cooperativity,janssen2014relaxation,janssen2016generalized}
GMCT thus provides a promising first-principles-based framework to extend the
applicability range of MCT-like approaches in both a qualitative and
quantitative manner.

The microscopic (i.e.\ fully wavevector-dependent) GMCT calculations reported
thus far have focused mainly on the dynamics for a fixed set of (hard-sphere)
densities.  It is not yet clear, however, how the higher-order microscopic GMCT
dynamics for a structural glass-former will change \textit{relative} to the
predicted critical point, whether standard MCT's scaling laws are successfully preserved, and how the predicted fragility may vary with
increasing GMCT closure levels. Indeed, since each new level in the GMCT
hierarchy simultaneously affects both the quantitative relaxation dynamics at a
given density (or temperature) as well as the location of the critical point,
the predicted dynamics may change in a non-trivial manner after rescaling with
respect to the new critical glass transition density.  Such an analysis is
important to assess the qualitative and universal features of the theory,
including the existence and validity of GMCT scaling laws near the glass
transition and in the $\beta$- and $\alpha$-relaxation regimes. 

In this paper, we report a comprehensive study on the glassy dynamics of hard
spheres within the microscopic GMCT framework. To enable a strictly
first-principles-based analysis, we use the analytic static structure factor
for hard spheres obtained from the Percus-Yevick closure to the
Ornstein-Zernike equation as input.\cite{wertheim1963exact} We include up to
five additional levels in the GMCT hierarchy (i.e.\ closing at the level of
twelve-point correlations) and show that the predicted critical packing fraction 
manifestly converges.  Interestingly, we also find that for a fixed distance
\textit{relative} to the respective critical point, GMCT predicts a
non-trivial \textit{slower} relaxation behavior as the closure level increases,
which is visible both in $F(k,t)$ and in the increased non-ergodicity
parameters.  These findings must be contrasted with GMCT calculations at
\textit{absolute} values of the packing fraction,\cite{janssen2015microscopic} in which
case higher-order GMCT always yields faster structural relaxation.
Finally, we perform a detailed scaling analysis in both the $\beta$- and
$\alpha$-relaxation regimes, and find that the successful scaling laws of
standard MCT are fully preserved within higher-order GMCT; additionally, the
corresponding critical exponents are quantitatively improved.  In particular, the predicted
fragility, von Schweidler exponent, and Kohlrausch parameters for hard spheres are all in 
better agreement with numerical simulations as
the GMCT closure level increases.  In the accompanying paper, we present an
analytic derivation of these scaling laws for GMCT at arbitrary order.

\section{Theory}
 
We first recapitulate the microscopic GMCT equations of motion first derived in Ref.\ \onlinecite{janssen2015microscopic}.
The dynamical objects of interest are the normalized $2n$-point density correlation functions
$\phi_n(k_1,\hdots,k_n,t)$, defined as
\begin{equation}
\label{eq:phindef}
\phi_n(k_1,\hdots,k_n, t) = \frac{\langle \rho_{\bm{-k_1}}(0) \hdots \rho_{-\bm{k_n}}(0)
	\rho_{\bm{k_1}}(t) \hdots \rho_{\bm{k_n}}(t) \rangle}
{\langle \rho_{\bm{-k_1}}(0) \hdots \rho_{-\bm{k_n}}(0)  
	\rho_{\bm{k_1}}(0) \hdots \rho_{\bm{k_n}}(0) \rangle},
\end{equation} 
where $\rho_{\bm{k}}(t)$ is a collective density mode at wavevector $\bm{k}$ and time $t$,
the angle brackets denote an ensemble average, and the label $n$ ($n=1,\hdots,\infty$) specifies the level of the hierarchy.
Note that for $n=1$ we have $\phi_1(k,t)=F(k,t)/S(k)$.
In the overdamped limit, the GMCT equations read
\begin{gather} 
\nu_n\dot{\phi}_n(k_1,\hdots,k_n,t) + \Omega^2_n(k_1,\hdots,k_n)\phi_n(k_1,\hdots,k_n,t) 
\nonumber \\ 
+\int_0^t M_n(k_1,\hdots,k_n,u) \dot{\phi}_n(k_1,\hdots,k_n,t-u) du  = 0 \label{eq:GMCTphi_n}, 
\end{gather} 
where $\nu_n$ is an effective friction coefficient, 
and
\begin{equation} 
\label{eq:Omega2n}
\Omega^2_n(k_1,\hdots,k_n) = D_0 \left[ \frac{k_1^2}{S(k_1)} + \hdots + \frac{k_n^2}{S(k_n)} \right] 
\end{equation}
are the so-called bare frequencies with $D_0$ denoting the bare diffusion constant. 
For the memory functions we have
\begin{eqnarray}
\label{eq:Mn} 
M_n(k_1,\hdots,k_n,t) = \frac{\rho D_0}{16\pi^3} \sum_{i=1}^n \frac{\Omega^2_1(k_i)}{\Omega^2_n(k_1,\hdots,k_n)}
\nonumber \\
\times \int d\bm{q} |\tilde{V}_{\bm{q,k}_i-\bm{q}}|^2 
S(q) S(|\bm{k}_i-\bm{q}|)
\hphantom{XXXX}
\nonumber \\
\times \phi_{n+1}(q,|\bm{k}_1-\bm{q}\delta_{i,1}|,\hdots,|\bm{k}_n-\bm{q}\delta_{i,n}|,t), \nonumber \\
\end{eqnarray}
where $\rho$ is the total density, $\delta_{i,j}$ is the Kronecker delta
function, and $\tilde{V}_{\bm{q,k}_i-\bm{q}}$ are the static vertices that represent
wavevector-dependent coupling strengths for the higher-level correlations.
These vertices are defined as
\begin{equation}
\label{eq:V}
\tilde{V}_{\bm{q,k-q}} = 
({\hat{\bm{k}}} \cdot \bm{q}) c(q) + 
{\hat{\bm{k}}} \cdot (\bm{k-q}) c(|\bm{k-q}|),
\end{equation}
where $\hat{\bm{k}} = \bm{k}/k$ and $c(q)$ denotes the direct
correlation function,\cite{hansen1990theory} which is related to the static structure factor as $c(q)
\equiv [1-1/S(q)] / \rho$. It is important to note that the latter serves as the
\textit{only input} to the theory. Equations (\ref{eq:GMCTphi_n})--(\ref{eq:V}) have been derived by
assuming convolution and Gaussian factorization approximations for all static multi-point correlators,
and by neglecting so-called off-diagonal dynamic multi-point correlators.\cite{janssen2015microscopic} 
The initial conditions for Eq.\ (\ref{eq:GMCTphi_n}) are $\phi_n(k_1,\hdots,k_n,0)=1$ for all $n$.

In order to solve the GMCT equations for finite order, a closure is necessary for the last included level $N<\infty$. 
One choice is to approximate the last level $\phi_N$ in a self-consistent manner by the product of $\phi_{N-1}$ and
$\phi_1$.
To account for permutation invariance of all wavenumber arguments $\{k_1,\hdots,k_n\}$, we apply the closure
\begin{eqnarray}
\label{eq:closure_t} 
M_N(k_1,\hdots,k_N,t) =\frac{1}{N-1}\frac{1}{\Omega^2_N(k_1,\hdots,k_N)}\times
\nonumber \\
\sum_{i=1}^{N}
\Omega^2_{N-1}(\{k_j\}^{(N-1)}_{j\neq i})M_{N-1}(\{k_j\}^{(N-1)}_{j\neq i},t)\phi_1(k_i,t),
\end{eqnarray}
where $\{k_j\}^{(N-1)}_{j\neq i}$ represents the $N-1$ wavenumbers in
$\{k_1,\hdots,k_N\}$ except the $k_i$.  Following earlier
convention,\cite{janssen2016generalized} the above closure relation is referred
to as a mean-field (MF) closure and is denoted as MF-$N[(N-1)^11^1]$. 
Note that the simplest MF
closure is the standard MCT factorization $\phi_2\sim\phi_1^2$, i.e.\ 
MF-$2[1^2]$.
\footnote{The notation MF-$N[n_1^{m_1}n_2^{m_2}]$ implies a closure of the
form $\phi_N \sim \phi_{n_1}^{m_1}\phi_{n_2}^{m_2}$.}
Another kind of closure
is a truncation of the hierarchy such that $\phi_N(k_1,\hdots,k_N,t)=0$, which
is equivalent to setting $\phi_{N-1}=\exp(-t/\nu_N)$. This is referred to as an
exponential (EXP-$N$) closure. It has been established
numerically,\cite{janssen2015microscopic} and proven mathematically for at
least one family of schematic GMCT models,\cite{janssen2016generalized} that
mean-field and exponential closures provide an upper and lower bound for the
relaxation dynamics in the limit $N\rightarrow\infty$, respectively.

Equations (\ref{eq:GMCTphi_n}), (\ref{eq:Mn}), and (\ref{eq:closure_t}) define
a unique solution.\cite{BiezemansBScthesis}
The solution is regular in the sense that all $\phi_N(k_1,\hdots,k_N,t)$ depend
smoothly on the $\nu_n$ and $M_n$ in any finite time interval because the
variation of $S(k)$ is small when going from liquid to glass. Notably, no \textit{ad hoc} or phenomenological 
assumptions are made regarding the existence of a
singularity or glassy relaxation features in the dynamics. 

Using the Laplace transform $F(s)=\mathcal L(f(t))(s)=\int_{0}^{\infty}f(t)\mathrm{e}^{-st}dt$ and the final value theorem,
we can obtain the non-ergodicity parameters $f_n(k_1,\hdots,k_n)\equiv \lim_{t\to\infty}\phi_n(k_1,\hdots,k_n,t)$ 
via
\begin{gather}
\label{eq:GMCT_long}
\frac{f_n(k_1,\hdots,k_n)}{1-f_n(k_1,\hdots,k_n)}= \frac{m_n(k_1,\hdots,k_n)}{\Omega^2_n(k_1,\hdots,k_n)},
\end{gather}
where $m_n(k_1,\hdots,k_n) \equiv \lim_{t\rightarrow\infty}M_n(k_1,\hdots,k_n,t)$ represents the long-time limit of the memory function.
Analogous to Eq.\ (\ref{eq:closure_t}), we apply a MF closure of the form
\begin{eqnarray}
\label{eq:closure_long} 
m_N(k_1,\hdots,k_N) =\frac{1}{N-1}\frac{1}{\Omega^2_N(k_1,\hdots,k_N)}\times
\nonumber \\
\sum_{i=1}^{N}
\Omega^2_{N-1}(\{k_j\}^{(N-1)}_{j\neq i})m_{N-1}(\{k_j\}^{(N-1)}_{j\neq i})f_1(k_i).
\nonumber \\
\end{eqnarray}
Note that an exponentional closure will always yield $f_n(k_1,\hdots,k_n)=0$ for all $n$, since then $\lim_{t\rightarrow\infty}M_N(k_1,\hdots,k_n,t)=0$.
By solving Eqs.\ (\ref{eq:GMCT_long}) and (\ref{eq:closure_long})
iteratively we can thus obtain the long-time limit of the dynamic density correlation
functions. These non-ergodicity parameters serve as an order parameter for the glass transition: 
for liquids $f_n$ is $0$, while for a glassy state $f_n>0$. The lowest packing fraction at which
the ergodicity-breaking transition occurs is referred to as the critical point $\varphi^c$. 

We numerically solve the microscopic GMCT equations [Eqs.\ (\ref{eq:GMCTphi_n})--(\ref{eq:closure_long})] 
for monodisperse Percus-Yevick hard spheres using an equidistant wavenumber grid of 100 points ranging from $kd=0.2$ to 
$kd=39.8$, where $d$ is the
hard-sphere diameter. The wavevector-dependent integrals over $\bm{q}$ in the memory functions are approximated
as a double Riemann sum.\cite{franosch1997asymptotic} 
For the time-dependent integration we use the algorithm described by Fuchs \textit{et al.},\cite{fuchs1991comments} 
starting with a time step size of $\Delta t=10^{-6}$ that is subsequently doubled every 32 points. 
Following Ref.\ \onlinecite{janssen2015microscopic}, we assume $D_0=1$ and set the effective friction coefficients $\nu_n$ to 1 for all $n$.

\section{Results and Discussion}

\begin{figure}
	\epsfig{file=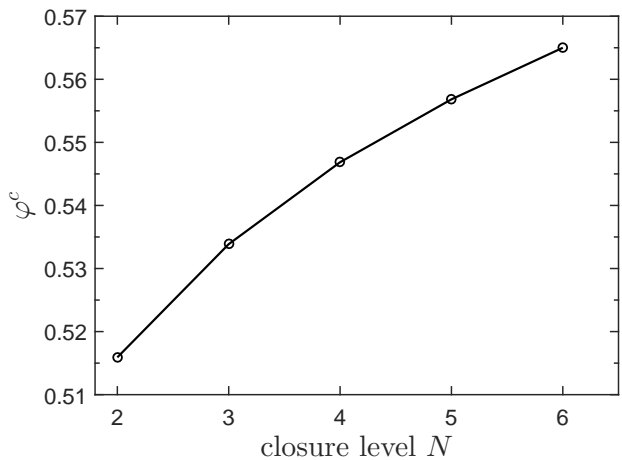,width=0.46\textwidth}
	\caption{\label{fig:phic_N} 
		The predicted critical packing fraction $\varphi^c$ for Percus-Yevick hard spheres 
                as a function of the GMCT closure level $N$, using mean-field closures of the form 
                MF-$N[(N-1)^11^1]$.
	}
\end{figure}
\subsection{Critical points and critical non-ergodicity parameters}

We first consider the GMCT solutions for the critical glass transition point.
It is well established that standard MCT, i.e.\ the lowest order GMCT MF
closure, predicts a glass transition for Percus-Yevick hard spheres at a
packing fraction of $\varphi^c=0.516$.\cite{franosch1997asymptotic} Figure
\ref{fig:phic_N} shows that, as the MF closure level increases from $N=2$ to
$N=6$, the critical point shifts toward higher values in a seemingly convergent (approximately logarithmic)
manner. This finding is fully consistent with the earlier $N=3$ and $N=4$
hard-sphere studies of Szamel and Wu and Cao, respectively, as well as with
infinite-order schematic GMCT models.\cite{mayer2006cooperativity,janssen2016generalized} We
emphasize that the convergence pattern is in fact far from
trivial, since there is no obvious small parameter in the theory.  At our
highest closure level studied, $N=6$, the predicted critical packing fraction
is 0.56(5), which is much closer to the experimental colloidal-hard-sphere
value\cite{van1991nonergodicity} of $\varphi_g=0.563$ than the standard MCT
prediction.  Based on the results of Fig.\ \ref{fig:phic_N}, we expect that the GMCT-predicted $\varphi^c$ will grow further beyond $N=6$, perhaps indefinitely, until the physical maximum of random close packing is reached ($\varphi\approx0.64$).\cite{brambilla2009probing} However, the Percus-Yevick
static structure factors that we use as input will become increasingly more
inaccurate at higher densities, and a fully quantitative comparison of our
results with experiment is therefore likely to break down above a certain
packing fraction. 
We also mention that in our current theory higher-order static correlation functions beyond $S(k)$ are not included as input, which may become more important at high densities and may lead to a faster convergence of $\varphi^c$ instead of a seemingly logarithmic convergence in Fig.~\ref{fig:phic_N}. 
Nonetheless, the uniformly convergent trend of $\varphi^c(N)$ in
Fig.\ \ref{fig:phic_N} is encouraging, as it suggests that a
\textit{finite}-order GMCT calculation--given the appropriate input
microstructure--may be sufficient to get an accurate first-principles
prediction of the precise location of the glass transition. 
 
\begin{figure}
	\epsfig{file=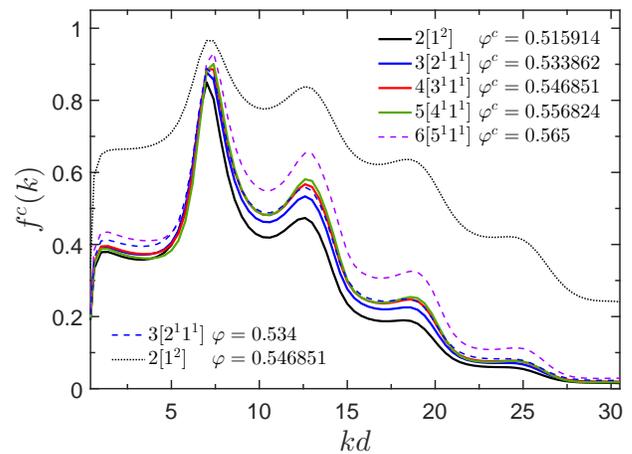,width=0.46\textwidth}
	\caption{\label{fig:fc} 
		The non-ergodicity parameters $f^c(k)$ as a function of wavenumber $k$ at the critical packing fraction 
                $\varphi^c$ for different GMCT MF closure levels. 
                Solid lines are the non-ergodicity parameters at the corresponding 6-digit critical packing fractions. 
                Dashed lines are the non-ergodicity parameters at the corresponding 3-digit critical packing fractions. 
                The black dotted line is the standard MCT prediction (MF-$2[1^2]$) for the long-time limit of the 
                two-point density correlation function at $\varphi=0.546851$, i.e.\ at 
                the critical packing fraction of MF-$4[3^11^1]$.
	}
\end{figure}

Let us now focus on the long-time limit of the two-point density correlation
functions at the critical points predicted by higher-order GMCT.  Figure
\ref{fig:fc} shows the critical non-ergodicity parameters $f^c(k)\equiv
f_1^c(k)$ for all MF closure levels $N$ considered in this work.  It can be
seen that increasing $N$ leads to overall higher non-ergodicity parameters at
all wavenumbers. Furthermore, the predicted $f^c(k)$ values manifestly converge
with $N$, at least for $N=2$ to $N=5$.  The values of $f^c(k)$ are, however,
very sensitive to the accuracy of the critical packing fraction. For example,
the blue dashed line in Fig.\ \ref{fig:fc} is obtained from the $N=3$
(MF-$3[2^11^1]$) closure using the critical packing fraction $\varphi^c=0.534$, i.e.\
$\varphi^c$ is determined up to 3 significant digits.  Under the same $N=3$
closure, however, a more precise prediction of $\varphi^c$ with 6 significant
digits, $\varphi^c=0.533862$, yields a markedly lower non-ergodicity parameter
(blue solid line in Fig.\ \ref{fig:fc}). In fact, the less accurate $N=3$
prediction is almost equivalent to the $N=4$ (MF-$4[3^11^1]$) non-ergodicity
parameter at the corresponding 6-digit critical packing fraction
$\varphi^c=0.546851$ (red solid line in Fig.\ \ref{fig:fc}).  For the highest
closure level considered, $N=6$, we could determine the critical point only up
to 3 significant digits within reasonable computing time; this prediction
constitutes an upper bound to the actual $\varphi^c$ at this level.  In view of
our lower-order GMCT results, we can therefore attribute the relatively large difference
in $f^c(k)$ between $N=5$ and $N=6$ to the inaccuracy of the critical packing
fraction for $N=6$, and we expect that the convergence pattern of $f^c(k)$ will
persist at all closure levels, provided that the numerical accuracy of
$\varphi^c$ is sufficiently large. 

At first glance, the overall increase of $f^c(k)$ with the GMCT closure level
$N$ may not seem suprising, since the respective critical packing
fractions--and thus the peak heights of the input $S(k)$--also increase with
$N$.  That is, a larger $f^c(k)$ might be regarded as a manifestation of a more
closely packed microstructure.  However, we argue that the increase of the
critical non-ergodicity parameter within higher-order GMCT is in fact a more
subtle and non-trivial effect.  Recall that the $N=3$ prediction for $f_1(k)$
at $\varphi=0.534$ is almost identical to the $N=4$ result at
$\varphi=0.546851$, even though the respective input structures are
significantly different.  Conversely, within standard MCT (i.e.\ GMCT with
$N=2$ closure), the MCT non-ergodicity parameters at $\varphi=0.534$ and  $\varphi=0.546851$ (dotted black line in Fig.~\ref{fig:fc})
will be significantly different,  
and both these standard-MCT results will be much \textit{higher} than the corresponding GMCT $N=3$ and $N=4$
$f_1(k)$ predictions. This suggests that there are two competing effects at
play in higher-order GMCT: on the one hand, increasing the GMCT closure level
will increase the critical packing fraction, consequently requiring
increasingly more peaked microstructures as input; on the other hand, for any
given packing fraction and input structure factor, a higher GMCT closure level
will lead to \textit{lower} non-ergodicity parameters.  The net outcome of
these two effects is an overall small increase in $f^c(k)$ that tends to
converge with $N$--a result that is \textit{a priori} far from trivial. We also note
that another means to disentangle the influence of the static structure factor
and the role of $N$ is to fix the static structure factor, i.e.\ assume a
density-independent $S(k)$, and then find the corresponding critical points to
obtain the new $f^c(k)$. This test also shows that even with identical input
structure factors, a higher GMCT closure level leads to a higher $f^c(k)$.
Hence, we conclude that the higher-order GMCT framework increases the critical
non-ergodicity parameters intrinsically, and not merely as a consequence of
using more closely packed microstructures as input.  This result, which
physically corresponds to relatively slower relaxation dynamics with $N$, will
also become apparent in the next section when considering the fully
time-dependent GMCT dynamics.

In view of the above new findings, let us also briefly revisit the earlier GMCT
studies by Szamel\cite{szamel2003colloidal} and Wu and Cao,\cite{wu2005high} who calculated the critical
non-ergodicity parameters for colloidal hard spheres up to $N=3$ and $N=4$ GMCT
closures, respectively. Their input static structure factors were obtained from
the Verlet-Weis correction to the Percus-Yevick expression.\cite{verlet1972equilibrium}
Both studies also found that an increase in $N$ leads to a larger $f^c(k)$, but
a quantitative comparison with experiment suggested that higher-order GMCT may
overestimate the magnitude of the non-ergodicity parameter. We point out,
however, that the critical packing fractions $\varphi^c$ reported in both
studies were determined only up to 3 significant digits, constituting a strict
and potentially large upper bound.  Indeed, our present work indicates that with
enhanced numerical accuracy of $\varphi^c$, the higher-order GMCT results for
$f^c(k)$ may decrease significantly and approach the experimental results more
closely.  Moreover, in the weakly polydisperse hard-sphere simulations of Ref.\
\onlinecite{weysser2010structural}, Weysser \textit{et al.}\ found that MCT underestimates the non-ergodicity parameters, which agrees with the trend we find here. They attributed this underestimation of $f^c(k)$ solely to the underestimation of $\varphi^c$, but we argue it should be a net outcome of the two effects discussed above. Future work should clarify whether the high-order GMCT
framework--given the appropriate microstructures as input--indeed correctly
converges upon the simulation and experimental data.

\subsection{Time-dependent relaxation dynamics}

\begin{figure*}
	\epsfig{file=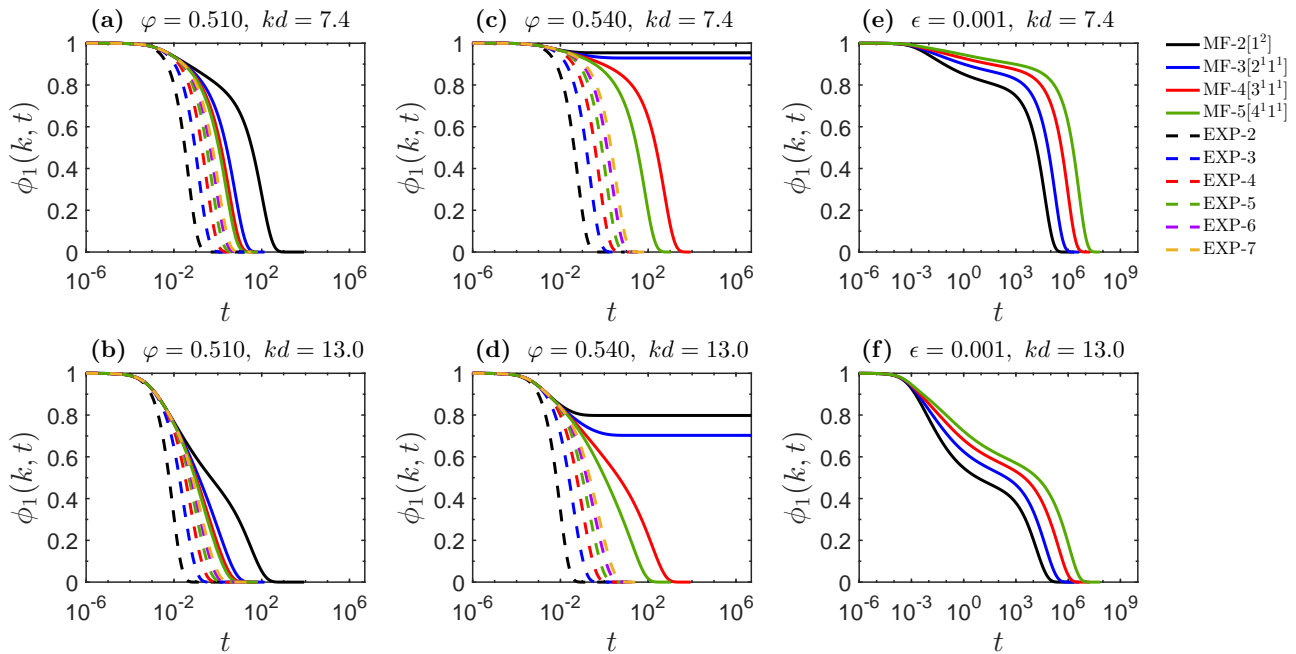,width=1.0\textwidth}
	\caption{\label{fig:Fkt} 
		Two-point density correlation functions $\phi_1(k,t)$ for Percus Yevick hard spheres under different GMCT closures. 
\textbf{(a)} $\varphi=0.510$ and $kd=7.4$; 
\textbf{(b)} $\varphi=0.510$ and $kd=13.0$; 
\textbf{(c)} $\varphi=0.540$ and $kd=7.4$; 
\textbf{(d)} $\varphi=0.540$ and $kd=13.0$; 
\textbf{(e)} $\epsilon=0.001$ and $kd=7.4$; 
\textbf{(f)} $\epsilon=0.001$ and $kd=13.0$. 
Solid lines correspond to GMCT MF-$N$ closures with $N=2,3,4,5$ and dashed lines to GMCT EXP-$N$ closures with $N=2,3,4,5,6,7$. 
The MF-$2[1^2]$ closure (black solid line) is equivalent to standard MCT.
	}
\end{figure*}
We now consider the explicit time-dependence of the dynamic density
correlators, obtained by solving Eq.\ (\ref{eq:GMCTphi_n}). The effect of
higher-order GMCT on the structural relaxation dynamics at a given packing
fraction has already been comprehensively studied in Ref.\
\onlinecite{janssen2015microscopic}, and we reiterate only the key conclusions here.
For all packing fractions and wavenumbers, MF closures provide an upper bound
to the dynamics while the exponential closures give a lower bound.  With
increasing closure level $N$, the two types of closures systematically converge
to each other. Figures \ref{fig:Fkt}(a) and (b) show the evolution of the
two-point density correlators $\phi_1(k,t)$ for Percus-Yevick hard spheres at
two different wavenumbers (corresponding to the first and second peak of
$S(k)$, respectively) at a packing fraction of $\varphi=0.510$. This value lies
below the standard-MCT critical point and the system is liquid for all closure levels. 
If we only consider the MF
closures, we can clearly see that for a fixed density the relaxation becomes
faster when increasing the closure level $N$. Since the exponential closures
approach the MF predictions from below, the EXP-$N$ series yields even more
liquid-like behavior than any MF solution.

Figures \ref{fig:Fkt}(c) and (d) 
show the time-dependent evolution of $\phi_1(k,t)$ at $\varphi=0.540$. This packing fraction lies above the predicted critical point
of both standard MCT and MF-$3$ GMCT. 
Indeed, the MF-$2[1^2]$ and MF-$3[2^11^1]$ curves for $\phi_1(k,t)$ do not decay to zero at any time scale 
(black and blue solid lines in \ref{fig:Fkt}(c) and (d)). 
When increasing the MF closure level from $N=3$ to $N=4$, it can be seen that the dynamics changes from glassy to liquid-like. 
This shift from non-ergodic to ergodic behavior is caused purely by the increase of the MF closure level $N$, since the
input microstructures are identical for all closure levels.
As before, the relaxation becomes faster when the MF closure level $N$ increases, 
and slower when the EXP closure level $N$ increases. 
Although the highest-order EXP and MF predictions are not as close to each other as in the low packing fraction case, 
the trend is exactly the same. Overall, the results of Fig.\
\ref{fig:Fkt}(a)-(d) confirm that the microscopic GMCT dynamics manifestly
converges with $N$, and that the inclusion of higher-order dynamic correlations
in GMCT at a fixed state point will reduce the degree of glassiness.  This also
corroborates our earlier conclusion that higher-order MF closures inherently
shift the glass transition point to larger densities. 

The results discussed so far concern the GMCT relaxation dynamics at an
\textit{absolute} value of the packing fraction.  To study the physical nature
of the glass transition, however, it is more instructive to consider the
dynamics close to and \textit{relative} to the critical point. Indeed, if the
predicted dynamics for all MF-$N$ closure levels would be identical at a fixed
\textit{relative distance} away from the respective critical points, then one
could conclude that higher-order GMCT only leads to a shift in the critical
density, rather than introducing a fundamentally new type of relaxation
dynamics.  To establish whether this is the case, we compare the time-dependent
density correlators $\phi_1(k,t)$ from different MF-$N$ closure levels at the
same reduced packing fraction
$\epsilon\equiv\frac{\varphi^c-\varphi}{\varphi^c}$. The parameter $\epsilon$
thus quantifies the relative distance to the $N$-dependent glass transition
points $\varphi^c$.  Figures \ref{fig:Fkt}(e) and (f) show this comparison for
a representative example of $\epsilon=0.001$ at the same wavenumbers as in
panels (a) and (b), respectively.  Interestingly, we find that in all cases the
structural relaxation dynamics at a fixed \textit{reduced} packing fraction
becomes \textit{slower} as the closure level $N$ increases, which is strikingly
opposite to the effect of increasing $N$ at a fixed \textit{absolute} packing
fraction [Fig.\ \ref{fig:Fkt}(a)-(d)].  
Furthermore, it can be seen that the
plateauing value of $\phi_1(k,t)$ in the $\beta$-relaxation regime [i.e.\ the
$t\sim10^0-10^3$ regime in Fig.\ \ref{fig:Fkt}(e) and (f)] also grows with
higher closure levels. In the limit of $\epsilon \rightarrow 0$, these
plateauing values will become exactly the non-ergodicity parameters $f^c(k)$,
thus rendering our explicitly time-dependent results fully consistent with the
data in Fig.\ \ref{fig:fc}. 

Overall, we can conclude that the incorporation of increasingly many GMCT
levels does not merely amount to a shift in the critical glass transition
point, but in fact leads to an inherent change in the structural relaxation
dynamics. More specifically, a higher closure level yields relatively more
glassy behavior, which is manifested as slower dynamics and a higher plateau in
the decay pattern of $\phi_1(k,t)$ at a given relative distance from the glass
transition singularity. We reiterate that, as discussed above in relation to
the critical non-ergodicity parameters, this effect cannot be simply attributed
to the underlying microstructure, but rather stems from the increased
complexity of the higher-order GMCT equations.  

\begin{figure}
	\epsfig{file=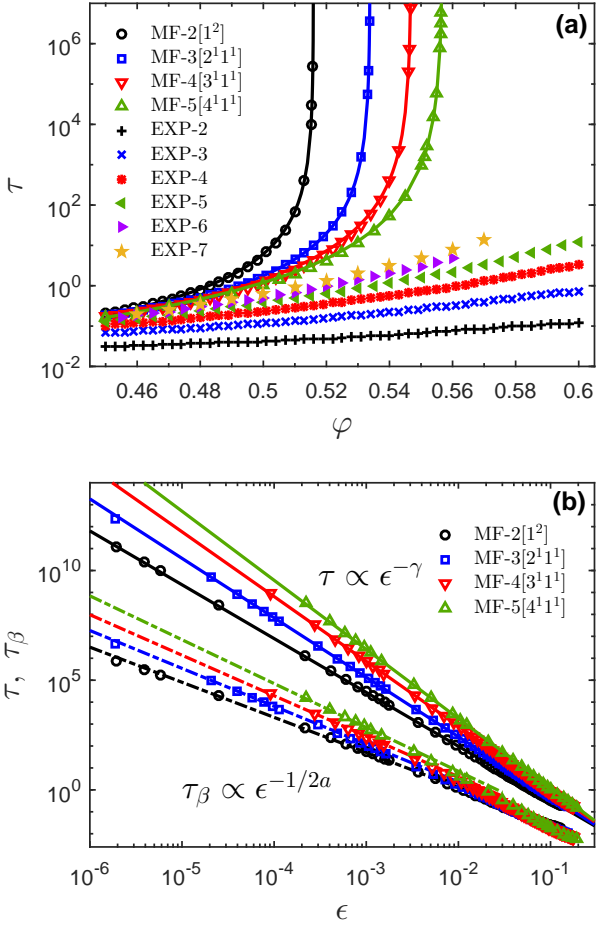,width=0.48\textwidth}
	\caption{\label{fig:tau} 
		Relaxation times at $kd=7.4$ for Percus-Yevick hard spheres using different GMCT closures. 
                \textbf{(a)} The $\alpha$-relaxation time $\tau$ as a function of the packing fraction $\varphi$. 
		Open and filled symbols are the numerical $\alpha$-relaxation times from GMCT obtained under MF and EXP closures, respectively. 
                The solid lines are the fitted power-law curves $\tau=\tau_{0}((\phi^c-\phi)/\phi^c)^{-\gamma}$ for the corresponding MF closure levels $N$. 
                The parameters $\phi^c$, $\gamma$, and $\tau_0$ are all $N$-dependent. 
		\textbf{(b)} Relaxation times as a function of the reduced packing fraction $\epsilon$. 
                The solid lines are the fitted power-law curves for the $\alpha$-relaxation time under MF closures, $\tau=\tau_{0}\epsilon^{-\gamma}$, as in panel (a). 
                The dash-dotted lines are the fitted curves for the $\beta$-relaxation time scale, $\tau_{\beta}=\tau_{\beta0}\epsilon^{-1/2a}$. 
                Here $a$ and $\tau_{\beta0}$ are also $N$-dependent. Different colors correspond to different closure levels.
	}
\end{figure}

\subsection{Scaling laws}

We now test the validity of several general scaling laws in the GMCT-predicted
glassy relaxation dynamics of Percus-Yevick hard spheres.  As already noted in
the introduction, it is firmly established that standard MCT makes several
universal and often remarkably accurate predictions on the scaling behavior of
$\phi_1(k,t)$. Specifically, these include:\cite{franosch1997asymptotic,gotze2008complex,reichman2005mode}
\begin{enumerate}
\item A power-law divergence of the $\alpha$-relaxation
time $\tau$ upon approaching the glass transition;
\item Different power laws associated with the onset and decay of the
$\beta$-relaxation regime, the exponents of which are related in a non-trivial
manner to the divergence of $\tau$;
\item A time-temperature (or time-density) superposition principle and stretched exponential relaxation for the
time-dependent decay of $\phi_1(k,t)$ in the $\alpha$-relaxation regime. 
\end{enumerate}
In the
accompanying paper, we show analytically that these scaling laws are fully
preserved within GMCT at arbitrary order under mean-field closures; in the following, we 
numerically test and extract the corresponding
critical exponents for the Percus-Yevick system as a function of the closure
level $N$.  It will be shown that the values of the exponents manifestly
converge with $N$, and that they are in good agreement with numerical
simulations of dense hard spheres.

\subsubsection{$\alpha$-relaxation time and fragility}

Figures \ref{fig:tau}(a) and (b) show the $\alpha$-relaxation times $\tau$
extracted from the GMCT solutions as a function of the absolute packing
fraction $\varphi$ and the reduced packing fraction $\epsilon$, respectively.
Here $\tau$ is defined via $\phi_1(k,\tau)=\mathrm{e}^{-1}$, with the wavenumber
$kd=7.4$ corresponding to the main peak of the static structure factor.  From
Fig.\ \ref{fig:tau}(a) we can observe that for a given $\varphi$, the
relaxation time decreases with increasing MF-$N$ closures, whereas the
relaxation time progressively increases for the EXP-$N$ closures. The
differences between the two types of closures become smaller as $N$ increases,
again establishing the uniform convergence of the GMCT hierarchy. In fact, at
low densities, e.g.\ $\varphi=0.47$, our highest-order GMCT predictions are
fully converged. These results are also consistent with the earlier microscopic
GMCT calculations for weakly polydisperse hard
spheres,\cite{janssen2015microscopic} and confirm that the inclusion of more
multi-point correlations can bring the system more deeply into the  density regime $\varphi>\varphi^c_{\text{MCT}}$ which is usually regarded as the activated glassy regime inaccessible for standard MCT. 

We find that the mean-field closure predictions for
$\tau(\varphi)$ conform to a functional form that is similar to the well-known
standard-MCT divergence.  Specifically, for all closure levels $N$ considered
in this work, the mean-field solutions are well described by a power law of the
form $\tau(\varphi)=\tau_{0}\epsilon^{-\gamma}$, where $\tau_{0}$ and
$\gamma$ are $N$-dependent fit parameters. 
These power-law fits are plotted as solid curves in Fig.\ \ref{fig:tau}(a) and
as straight solid lines in Fig.\ \ref{fig:tau}(b).  Importantly, while the
\textit{qualitative} power-law form remains the same at least up to $N=5$, 
the
power-law exponent $\gamma$ increases \textit{quantitatively} with the
mean-field closure level.  Table \ref{tab:coeff} lists the explicit
$N$-dependent values of $\gamma$; note that for $N=3$ and $N=4$ we have
included two different types of mean-field closures.  
It can be seen that $\gamma$ changes significantly and monotonically over the full range of MF-$N$ closures
considered, increasing by $28\%$ when going from level $N=2$ to $N=5$. 
Furthermore, the fitted value of $\gamma$ grows in a seemingly convergent manner with $N$, 
akin to the convergence pattern of $\varphi^c$.

The $N$-dependent increase of $\gamma$ also carries an important physical
implication that was already anticipated in an earlier schematic infinite-order
study:\cite{janssen2014relaxation} the \textit{fragility} predicted by higher-order
GMCT can be significantly different from the standard MCT result--even when the same static structure factors are used as input.
For the
Percus-Yevick system considered in this work, increasing the mean-field closure
level up to $N=5$ keeps the material fragile, but a higher $N$ leads to a lower fragility
index, i.e.\ a more gradual vitrification process as compared to standard MCT. Note that this trend is general, regardless of the packing fraction $\varphi_g$ at which the fragility index would be defined.
Although no experimental or simulation data are available for the dynamics of
Percus-Yevick hard spheres, it is reasonable to compare our theoretical
predictions against the weakly polydisperse hard-sphere simulations of Ref.\
\onlinecite{weysser2010structural}. The extracted $\gamma$ exponent for 3-component
hard spheres was found to be $2.63$,\cite{weysser2010structural} and careful
inspection of the data (Fig.\ 23 of Ref.\ \onlinecite{weysser2010structural})
suggests that $\gamma$ may be as large as $2.9$ when considering only small
values of $\epsilon$, i.e.\ densities very close to the glass transition point.
First-principles standard MCT calculations for the 3-component system predicted
a $\gamma$ value of 2.445,\cite{weysser2010structural} implying that standard
MCT overestimates the fragility index. Our higher-order GMCT framework is capable of 
remedying this problem, and indeed our MF-$N$ predictions for $N>2$ approach
the empirical $\gamma$ parameter more closely. Overall, this suggests that
microscopic GMCT can provide a more quantitatively accurate first-principles
prediction for the fragility index of fragile structural glass-formers. We also note
that experimental data on soft colloids suggest that a decreased fragility is accompanied by an 
increased non-ergodicity parameter (see Fig.\ 1 of Ref.\ \onlinecite{mattsson2009soft}); 
this trend is also fully consistent with the here reported high-order GMCT results. 

For the exponential closure series, the $\alpha$-relaxation times of the Percus-Yevick system are always lower than those 
of the corresponding MF closure series. As the EXP-$N$ closure level increases, the relaxation time also 
increases and becomes closer to the MF solutions. However, in contrast to the MF-$N$ series,
the EXP-$N$ results for $N\leq 7$ show only a modest growth of the relaxation time, predicting
an increase of $\tau$ of only two orders of magnitude over the full 
range of packing fractions considered. Consistent with earlier work,\cite{janssen2015microscopic} this implies 
that the emergence of strongly glassy behavior from an EXP-$N$ GMCT closure will require relatively
large values of $N$.

The shape of $\tau(\varphi)$ under EXP-$N$ GMCT closures cannot be easily captured in an analytic formula 
and we find that the shape is also sensitive to the selected wavenumber. 
However, for the wavenumber considered here, $kd=7.4$, 
the EXP-$N$ results for $\tau(\varphi)$ appear consistent with the \textit{onset} of a power law that diverges at very large 
 $\varphi^c$ and that may possibly cross over into an avoided transition.\cite{brambilla2009probing} Also note that the predicted $\tau(\varphi)$ values 
tend to deviate more clearly from an Arrhenius curve as the EXP-$N$ closure level increases. This implies that, for the hard-sphere
system presently under study, the predicted fragility should change from strong to more fragile upon increasing the exponential closure level.

Although computational limitations currently prevent us from calculating closure levels beyond $N=7$,
let us briefly remark on the scenario that may emerge from GMCT in
the limit of $N\rightarrow\infty$. In general, the 
EXP-$N$ and MF-$N$ series should fully converge in this limit, 
ultimately yielding a unique $\tau(\varphi)$ solution that is insensitive to the
details of the closure approximation.\cite{mayer2006cooperativity,janssen2016generalized} 
In the normal liquid regime, we can expect 
the EXP-$N$ series to converge relatively fast with $N$, whereas close to the glass transition 
the self-consistent MF-$N$ closures will likely perform better.\cite{janssen2015microscopic} 
For dense hard spheres, the exact solution
should conform to a fragile growth behavior; the here presented MF-$N$ solutions already
predict a fragile pattern for all $N$ levels considered, and we may expect that our EXP-$N$ solutions  
will eventually cross over into a marked non-Arrhenius form for sufficiently large $N$.

\begingroup
\setlength{\tabcolsep}{10pt} 
\renewcommand{\arraystretch}{1.1} %
\begin{table}
	\caption{Predicted critical packing fractions $\varphi^c$ and parameters $\gamma$, $a$, $b$, $\lambda$ for Percus-Yevick hard spheres obtained 
         under different GMCT MF-$N$ closures.
	    }
	\begin{tabular}{lccccc} 
	\hline
	\hline
\\[-1em]
	MF level & $\varphi^c$&$\gamma$&$a$&$b$&$\lambda$\\
	\\[-1em]
	\hline
	\\[-1em]
	$2[1^2]$ &0.515914 &2.46&0.31&0.59&0.73\\
	$3[1^3]$ & 0.526624 &2.58&0.30&0.55&0.76\\
	$3[2^11^1]$ & 0.533862 &2.71&0.29&0.51&0.78 \\
	$4[1^4]$ & 0.535382 &2.71&0.29&0.51&0.78\\
	$4[3^11^1]$ & 0.546851 &2.95&0.27&0.45&0.81\\
	$5[4^11^1]$  & 0.556824 &3.15&0.25&0.43&0.83\\
	\\[-1em]
\hline
\hline
	
	\end{tabular}

	\label{tab:coeff}
\end{table}
\endgroup

\subsubsection{$\beta$-relaxation regime}
To study the GMCT scaling laws in the $\beta$-relaxation regime, i.e.\ the
intermediate-time behavior of $\phi_1(k,t)$ associated with the cage effect,
let us first recapitulate the general predictions of standard MCT in this
domain.\cite{franosch1997asymptotic,gotze2008complex,kob2002supercooled,leutheusser1984dynamical}
\begin{itemize}
\item The $\beta$-relaxation regime can be characterized by a unique time
scale $\tau_{\beta}$ (also sometimes denoted as $\tau_{\sigma}$
\cite{franosch1997asymptotic}), which is defined as
$\phi_1(k,\tau_{\beta})=f^c(k)$. Standard MCT predicts that $\tau_{\beta}$
conforms to a power law of the form $\tau_{\beta}\sim\epsilon^{-1/2a}$, where
$a$ is a constant.
\item The $\beta$-relaxation of $\phi_1(k,t)$ is
predicted to obey a time-wavenumber factorization property such that
$\phi_1(k,t) = f^c(k)+ h(k) G(t)$, where the time-independent function $h(k)$ represents the so-called critical amplitude. 
\item Asymptotically close to the critical point, $\epsilon\approx0$, the time-dependent 
onset to and decay away from the plateau are described to leading order by 
\begin{numcases}{G(t) \sim}
 t^{-a} & $\text{if }t < \tau_{\beta}$, \nonumber \\
 t^{b}  & $\text{if }t > \tau_{\beta}$, \nonumber 
\end{numcases}
where $a$ is the same constant as above.
\item Further away from the critical point, the $k$-independent function $G(t)$ scales
with the reduced packing fraction $\epsilon$ as $G(t)= \sqrt{\epsilon}g_{\pm}(t/\tau_{\beta})$ where $g_+(t/\tau_{\beta})\sim(t/\tau_{\beta})^{-a}$ and
$g_-(t/\tau_{\beta})\sim(t/\tau_{\beta})^{b}$. The functional forms of $g_{+}(t/\tau_{\beta})$ and $g_{-}(t/\tau_{\beta})$ are also known as the critical decay and the von Schweidler law, respectively.\cite{kob2002supercooled}\item MCT predicts that the parameters $a$ and
$b$ obey the relation
$\lambda=\Gamma(1-a)^2/\Gamma(1-2a)=\Gamma(1-b)^2/\Gamma(1-2b)$, and they are
also related to the power-law exponent $\gamma$ of the
$\alpha$-relaxation time as $\gamma=1/2a + 1/2b$. This thus points toward a non-trivial but intimate
connection between the early $\beta$, late $\beta$, and $\alpha$-relaxation processes.
\end{itemize}
As shown analytically in the
accompanying paper, these scaling laws can be generalized to the higher-order
GMCT framework under mean-field $N>2$ closures; interestingly, all MCT scaling
laws are rigorously preserved within GMCT for arbitrary closure levels $N$.
However, the parameters $a$, $b$, and $\lambda$, as well as the exponent
$\gamma$ discussed earlier, now become explicitly $N$-dependent. 

From a numerical point of view, the scaling behavior of $\tau_{\beta}$ as a
function of $\epsilon$ can be tested most easily, since $f^c(k)$ is
well-defined. Figure \ref{fig:tau}(b) shows our GMCT Percus-Yevick hard-sphere
predictions of $\tau_{\beta}$ at wavenumber $kd=7.4$ under different MF-$N$
closures.  It can be seen that $\tau_{\beta}$ indeed accurately conforms to the
power law $\tau_{\beta} \propto \epsilon^{-1/2a}$ for all values of $N$,
allowing us to directly extract the $N$-dependent parameters $a$ from the
fitted power-law exponents. The parameters $b$ then readily follow from the
generalized relation $\gamma=1/2a + 1/2b$, where $\gamma$ is the power-law
exponent of the $\alpha$-relaxation time.  It can subsequently be verified that
$a$ and $b$, within the same closure level $N$, also satisfy the analytic GMCT relation
$\lambda=\Gamma(1-a)^2/\Gamma(1-2a)=\Gamma(1-b)^2/\Gamma(1-2b)$, thus
confirming the robustness of our analysis.  The corresponding $a$, $b$,
$\lambda$ parameters for Percus-Yevick hard spheres under different MF closures 
are shown in Table~\ref{tab:coeff}.  Note that all parameters
monotonically change with $N$ and that they may converge for sufficiently large
$N$.

\begin{figure}
	\epsfig{file=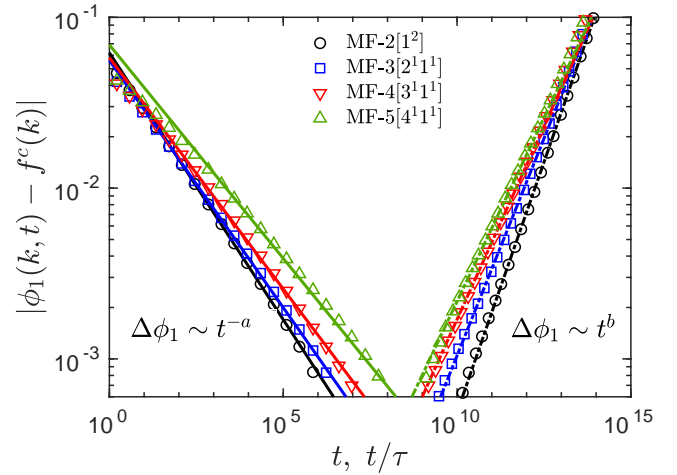,width=0.48\textwidth}
	\caption{\label{fig:dynamics_critical} 
		Relative two-point density correlation functions
$\Delta\phi_1(k,t)=|\phi_1(k,t)-f^c(k)|$ for Percus-Yevick hard spheres at $kd=7.4$ and $\epsilon\approx0$. The
packing fractions $\varphi$ are $0.515913$, $0.533861$, $0.546850$, $0.556823$ for MF-$N$
closure levels $N=2, 3, 4, 5$, respectively. 
The symbols represent the numerical GMCT critical dynamics for $|\phi_1(k,t)-f^c(k)|$. The solid and dashed
lines are fits of $\Delta\phi_1(k,t)\sim t ^{-a}$ and $\Delta\phi_1(k,t)\sim t^{b}$, respectively, using the 
corresponding $N$-dependent $a$ and $b$ exponents of Table \ref{tab:coeff}. For clarity, we rescale the time of the late $\beta$-relaxation regime by the corresponding $\alpha$-relaxation time $\tau$ and shift the lines horizontally by a factor $10^{15}$, i.e. $t/\tau\times 10^{15}$.
Note that the obtained $f^c(k)$ is an upper bound for the non-ergodicity parameters, and that $\Delta\phi_1(k,t)$ becomes very sensitive to numerical noise when $\phi_1(k,t)$ is very close to  $f^c(k)$. For the lowest values shown here ($<5\times10^{-3}$), we have therefore used $\Delta\phi_1(k,t)\approx|\phi_1(k,t)-0.999f^c(k)|$.	}
\end{figure}

\begin{figure}
	\epsfig{file=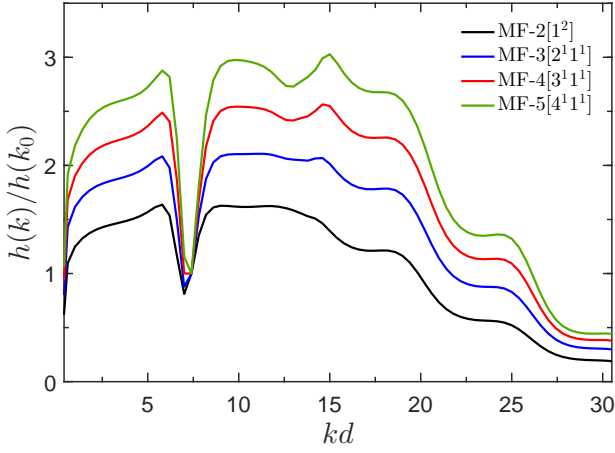,width=0.46\textwidth}
	\caption{\label{fig:Yq} 
		Rescaled critical amplitudes $h(k)/h(k_0)$ from the critical
 dynamics where $k_0d=7.4$. The packing fractions are $0.515913$, $0.533861$,
 $0.546850$ and $0.556823$ for MF-$N$ closure levels $N=2,3,4,5$, respectively.
	}
\end{figure}

Next we study the full dynamics in the $\beta$-relaxation regime. Here we only
focus on the behavior of $\phi_1(k,t)$; the scaling laws for arbitrary
$\phi_n(k_1,\hdots,k_n,t)$ with $n\leq N$ are derived in the accompanying paper. 
We first test the critical decay $\phi_1(k,t)- f^c(k)=h(k)(t/t_0)^{-a}\sim t^{-a}$ (with $t_0$ a fit parameter) 
and the von Schweidler law $\phi_1(k,t)-f^c(k)=-h(k)(t/\tau)^{b}\sim
t^{b}$ in the early and late $\beta$-relaxation regime, respectively, when
$\epsilon\approx0$.
We emphasize that $a$, $b$, $h(k)$, $t_0$, $\tau$ and $f^c(k)$
are all explicitly dependent on the mean-field closure level $N$.  
In Fig.\ \ref{fig:dynamics_critical} we plot
the relative correlation functions $\Delta\phi_1(k,t)=|\phi_1(k,t)-f^c(k)|$ at
wavenumber $kd=7.4$ as obtained from our numerical GMCT calculations, as well as the fitted functions 
using the $N$-dependent $a$ and $b$ parameters obtained from the above procedure. 
The packing fractions used for this analysis are all marginally below 
the respective MF-$N$ critical points to ensure an adequate probing of the critical dynamics. 
Figure \ref{fig:dynamics_critical} shows that both power laws are
in good agreement with our numerical GMCT predictions of $\phi_1(k,t)$ for all
values of $N$ considered. It must be emphasized that the critical exponents $a$ and $b$ are not used as
free fit parameters here, but rather follow from the earlier scaling analysis of $\tau_{\beta}$.
Note that $\Delta \phi_1(k,t)$ becomes very sensitive to
$f^c(k)$ when $\Delta \phi_1(k,t)$ is close to $f^c(k)$. We therefore only show and
fit the data when $\Delta \phi_1(k,t)$ is higher than $~5\times10^{-4}$.  
 We can also extract the critical amplitude $h(k)$ from the critical dynamics. A
widely used method is selecting two different time $t_1$ and $t_2$ in the
$\beta$-relaxation regime and calculating
\cite{weysser2010structural,gleim2000relaxation} 
\begin{equation}
\frac{h(k)}{h(k_0)}=\frac{\phi_1(k,t_1)-\phi_1(k,t_2)}{\phi_1(k_0,t_1)-\phi_1(k_0,t_2)}.
\end{equation} 
Figure~\ref{fig:Yq} shows $h(k)/h(k_0)$ with $k_0d=7.4$ for
different closure levels using this method. In principle, the $N$-dependent
$h(k)$ can also be analytically derived; this derivation is discussed in the
accompanying paper.

We now turn to the scaling laws in the $\beta$-relaxation regime for $\epsilon>0$. Our higher-order GMCT framework predicts
that, to leading order, the asymptotic behavior of the $2$-point dynamic density correlators $\phi_1(k,t)$ obeys a 
factorization relation similar to that in standard MCT:
\begin{gather}
\phi_1(k,t)-f^c(k)
=h(k)G(t),
\label{eq:betacritical}
\end{gather}
with
\begin{equation}
\label{eq:G}
G(t)=\sqrt{\epsilon}g_{\pm}(t/\tau_{\beta}),
\end{equation}
where $g_+(t/\tau_{\beta})\sim(t/\tau_{\beta})^{-a}$ and
$g_-(t/\tau_{\beta})\sim(t/\tau_{\beta})^{b}$.  This relation applies for all
MF-$N$ closures. Unlike standard MCT, however, the functions
$g_{\pm}(t/\tau_\beta)$ depend explicitly on the closure level $N$, owing to
the $N$-dependence of the exponents $a$ and $b$. Also, the non-ergodicity
parameters $f^c(k)$ and the critical amplitude $h(k)$ depend on the closure level $N$
as mentioned before.

\begin{figure}
	\epsfig{file=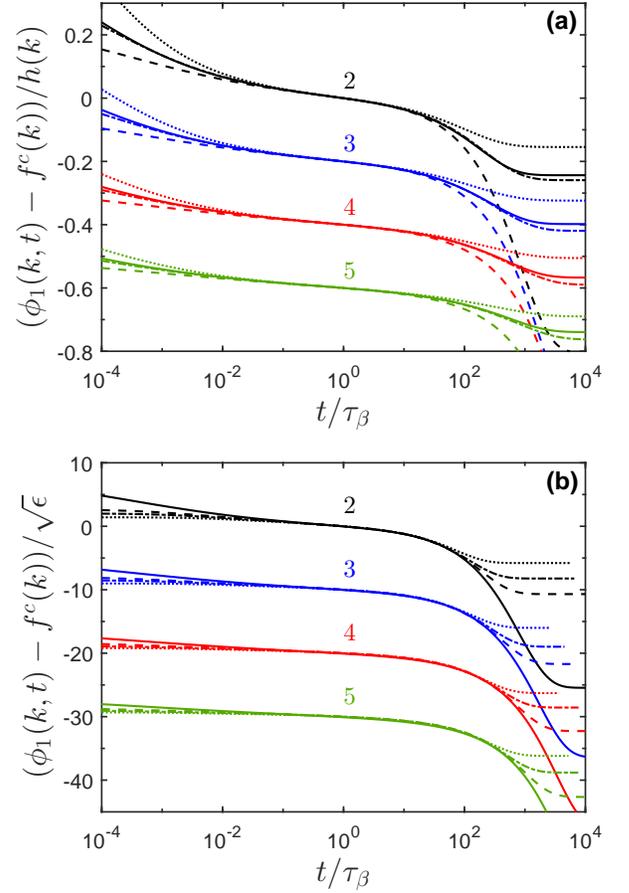,width=0.46\textwidth}
	\caption{\label{fig:beta} 
		$\beta$-relaxation scaling laws for different MF closure levels. 
		\textbf{(a)} Scaling with the critical amplitude $h(k)$ [Eq.\ (\ref{eq:betacritical})] at $\epsilon\approx0.001$. 
                 The relative correlation functions are normalized by the corresponding $h(k)$ at four different wavenumbers: 
                 $kd=3.4$ (solid lines), $kd=7.4$ (dashed lines), $kd=10.6$ (dash-dotted lines), and $kd=17.4$ (dotted lines). 
                 For clarity, the lines are shifted by a factor of $0.2\times(N-2)$ for every level $N$.
		\textbf{(b)} Scaling with $\epsilon$ [Eq.\ (\ref{eq:G})] for wavenumber $kd=7.4$. 
                The relative correlation functions are scaled by the corresponding $1/\sqrt{\epsilon}$ at four different $\epsilon$ values:
                $\epsilon=0.001$ (solid lines), $\epsilon=0.005$ (dashed lines), $\epsilon=0.01$ (dash-dotted lines), and 
                $\epsilon=0.02$ (dotted lines). For clarity, the lines are shifted by a factor of $10\times(N-2)$ for every level $N$.
	}
\end{figure}

With the critical amplitudes shown in Fig.~\ref{fig:Yq} we can test the scaling
of $\phi_1(k,t)$ with wavenumber $k$. Figure \ref{fig:beta}(a) shows the scaled
relative density correlation functions for different wavenumbers and for
different MF-$N$ closures at $\epsilon\approx0.001$. For a given closure level
$N$ and reduced packing fraction $\epsilon$, all curves with different wavenumbers collapse to one curve
around $t=\tau_{\beta}$, which is $\sim g_{\pm}(t/\tau_\beta)$. This
confirms the predicted scaling with $h(k)$ in Eq.~(\ref{eq:betacritical}).

We complete our $\beta$-relaxation analysis by testing the scaling of $G(t)$
with $\epsilon$. Figure \ref{fig:beta}(b) shows the data collapse for
$\phi_1(k,t)$ according to Eqs.\ (\ref{eq:betacritical})--(\ref{eq:G}) for
different values of $\epsilon$ and for different MF-$N$ closures. We here present
only the Percus-Yevick results at wavenumber $kd=7.4$; we have verified that for other
wavenumbers the results are similar.  It may be seen that all curves fully collapse
at $t=\tau_{\beta}$, indicating that the predicted square-root scaling with
$\epsilon$ in $g_{\pm}(t)$ is fully preserved in our numerical higher-order
GMCT calculations. 

The general scaling laws in the $\beta$-relaxation regime are among the biggest triumphs of 
standard MCT. Indeed, many experiments and simulations on supercooled liquids
show e.g.\ a time-wavenumber factorization and a data collapse with $g_{\pm}(t/\tau_\beta)$. 
However, the MCT-predicted exponents $a$ and $b$
are usually not quantitatively accurate.\cite{voigtmann2004tagged}
The detailed study for weakly polydisperse hard spheres by Weysser \textit{et al.}\cite{weysser2010structural} 
found that standard MCT overestimates the value of $b$, even when accounting for the polydispersity through a multi-component MCT analysis.
The here presented $b$ exponents for Percus-Yevick hard spheres are found to decrease in a convergent manner with increasing closure level $N$ 
(Table \ref{tab:coeff}), suggesting that it will be fruitful to extend the work of Weysser \textit{et al.}\cite{weysser2010structural} to higher-order GMCT.
Furthermore, in simulations of a binary
Lennard-Jones mixture,\cite{nauroth1997quantitative,kob1995testing} which is
also a fragile system, the authors obtained the $\lambda$ parameter from both the $\beta$
regime analysis, i.e.\ by fitting the density correlators with the von
Schweidler law to obtain the exponent $b$, and the theoretical standard MCT
calculation. The fitted $\lambda$ was found to be $0.78\pm0.02$, which is larger than the
MCT prediction of $0.708$. This result also agrees with the trend of $\lambda$ we find here:
as can be seen 
from Table \ref{tab:coeff}, the higher the GMCT closure level $N$, the
larger the value of $\lambda$. These findings suggest that GMCT indeed can provide a more quantitatively accurate prediction of the critical
exponents, but more work is needed to firmly establish the accuracy of the theory for a realistic glass-forming material.

\subsubsection{$\alpha$-relaxation regime: time-density superposition principle and stretched exponential decay}

Lastly, we return to the $\alpha$-relaxation regime and 
test the existence of a time-temperature or time-density superposition principle and stretched exponential decay within microscopic GMCT.
The superposition principle states that the final decay of $\phi_1(k,t)$ can be collapsed onto a temperature- or 
density-\textit{independent} master function such that 
$\phi_1(k,t) = \tilde{\phi}_1(k,t/\tau)$.
That is, after absorbing all explicit temperature- and density-dependence into the $\alpha$-relaxation time, and by subsequently rescaling the time with $\tau$,
$\phi_1(k,t)$ conforms to a single master curve at all temperatures and densities. 
It has been shown that this superposition principle is obeyed within standard MCT,\cite{gotze2008complex,franosch1997asymptotic,kob2002supercooled} but we find that it also generally applies
within higher-order GMCT under arbitrary mean-field closure levels. 
To order $\sqrt{\epsilon}$, the GMCT-predicted $\alpha$-relaxation of all $\phi_1(k,t)$ correlators satisfies the relation
\begin{equation}
\phi_1(k,t) = \tilde{\phi}_1(k,t/\tau)=f^c(k)-h(k)(t/\tau)^b.
\label{eq:alphascaling}
\end{equation}

\begin{figure}
	\epsfig{file=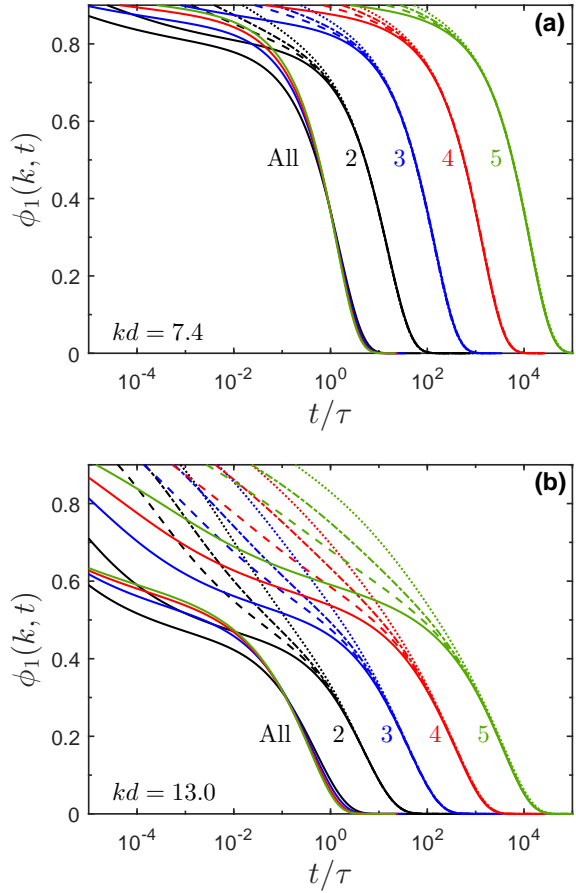,width=0.46\textwidth}
	\caption{\label{fig:alpha} 
		$\alpha$-relaxation scaling laws for different MF closure levels.
		\textbf{(a)} Two-point density correlation functions for wavenumber $kd=7.4$ at four different $\epsilon$ values: 
                $\epsilon=0.001$ (solid lines), $\epsilon=0.005$ (dashed lines), $\epsilon=0.01$ (dash-dotted lines), and $\epsilon=0.02$ (dotted lines). 
                Different colors correspond to different MF closure levels $N$. For clarity, the lines are shifted horizontally by a factor of $10^{(N-1)}$ for each level $N$.
		The solid curves labeled as 'All' are all MF-$N$ results ($N=2,3,4,5$) for $\epsilon\approx0.001$. 
		\textbf{(b)} Same as (a) except that $kd=13.0$. We point out that for a given closure level $N$ and $\epsilon$, 
                the $\tau$ used here is exactly the same as in (a), i.e.\ the $\alpha$-relaxation time at $kd=7.4$ [defined as $\phi_1(kd=7.4,\tau)=\mathrm{e}^{-1}$]. }
\end{figure}

In Fig.\ \ref{fig:alpha}(a) and (b) we test the collapse of $\phi_1(k,t)$ onto $\tilde{\phi}_1(k,t/\tau)$ for all MF-$N$ closures
and $\epsilon$ values considered in this work. We consider two different wavenumbers, $kd=7.4$ (Fig.~\ref{fig:alpha}(a)) and $kd=13.0$ (Fig.~\ref{fig:alpha}(b)). 
These results show that all curves collapse at a given closure level $N$ for different values of $\epsilon$. 
This numerically confirms that the superposition principle is rigorously 
obeyed in the $\alpha$-relaxation regime at all GMCT closure levels. 
Note that here for a given $N$ and $\epsilon$, we use the same $\tau(N,\epsilon,k_0d=7.4)$ for both wavenumbers; this further indicates that for all GMCT MF-$N$ levels 
the power law of the $\alpha$-relaxation time, $\tau\sim\epsilon^{-\gamma}$, is universal for all wavenumbers. 
As can be seen in Eq.~(\ref{eq:alphascaling}), 
the $\tilde{\phi}_1(k,t)$ also depends on the closure level $N$ because of the $N$-dependent $f^c(k)$, $b$ and $h(k)$. 
This can be seen by overlapping the $\tilde{\phi}_1(k,t)$  curves for different GMCT MF-$N$ closure levels, labeled by 'All' in Fig.~\ref{fig:alpha}(a). 
Even though all closures correspond to the same $\epsilon$ and wavenumber, the curves for different $N$ do not fully collapse. 
The difference is more pronounced at $kd=13.0$. Also note that Eq.~(\ref{eq:alphascaling}) describes the relaxation starting from the 
late $\beta$-relaxation regime. Considering that $\epsilon\sim\tau^{-1/\gamma}\sim\tau_{\beta}^{-2a}$, we can thus identify Eq.~(\ref{eq:alphascaling}) for the $\alpha$-relaxation regime 
with the $\beta$-relaxation dynamics of Eq.~(\ref{eq:betacritical}) and Eq.~(\ref{eq:G}) with $g_-$. 
This confirms that for all MF-$N$ levels these scaling laws are applicable as in standard MCT, but with different exponents.

\begin{figure}
	\epsfig{file=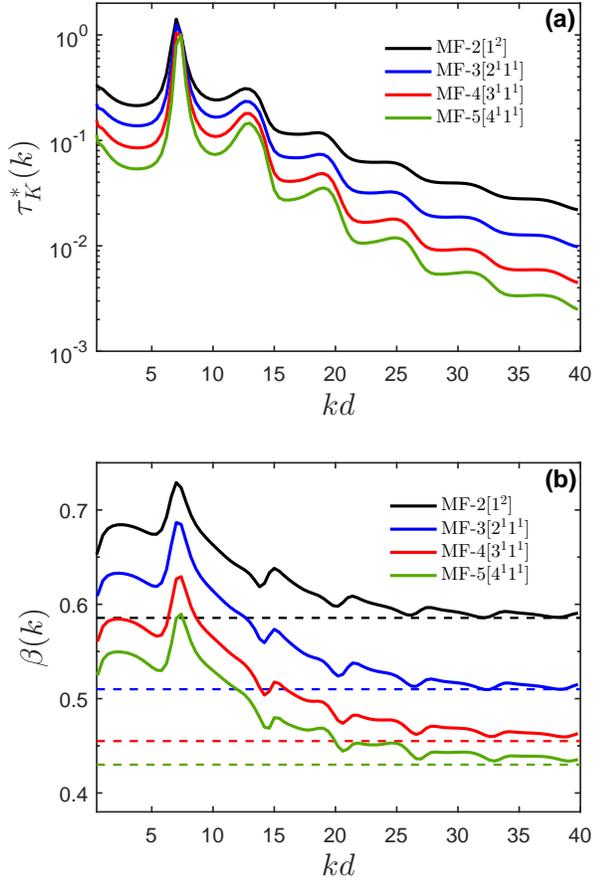,width=0.46\textwidth}
	\caption{\label{fig:Kohlrausch} 
		Fit parameters for the stretched-exponential Kohlrausch function in the $\alpha$-relaxation regime for different MF-$N$ closure levels. 
                All curves are obtained by fitting Eq.~(\ref{eq:Kohlrausch}) for all GMCT MF-$N$ levels at $\epsilon=0.001$. 
                The fit ranges are $t\in[4.3\times10^2:1.5\times10^4]$, $t\in[5.2\times10^2:5.3\times10^4$, $t\in[4.2\times10^3:1.2\times10^5]$,$t\in[7\times10^3:4\times10^5]$ 
                for levels $N=2, 3, 4, 5$, respectively. Within these ranges the parameters exibit only a weak dependence on the fit boundaries.
		\textbf{(a)} Rescaled $\alpha$-relaxation times $\tau^*_K(k)=\tau_K(k)/\tau_K(k_0)$ as a function of wavenumber $k$. 
		\textbf{(b)} Kohlrausch stretching exponents $\beta(k)$ as a function of wavenumber $k$. The dashed lines are the corresponding parameters $b$ of Table~\ref{tab:coeff}.}
\end{figure}

The final $\alpha$-relaxation process can be well described by the stretched-exponential Kohlrausch function 
\begin{equation}
\phi_1(k,t)=A(k)\exp{\left[-\left(\frac{t}{\tau_K(k)}\right)^{\beta(k)}\right]}.
\label{eq:Kohlrausch}
\end{equation}
In standard MCT, $A(k)\leq f^c(k)$ and when $k\rightarrow\infty$, $\beta(k)\rightarrow b$.\cite{fuchs1994kohlrausch} 
In the following we numerically demonstrate that this equation is also applicable for GMCT, but the $A(k)$, $\tau_K(k)$ and $\beta(k)$ 
depend on the MF closure level $N$. Figures~\ref{fig:Kohlrausch}(a) and (b) show the GMCT fit parameters $\tau_K(k)$ and $\beta(k)$ for Percus-Yevick hard spheres
at $\epsilon=0.001$.
Although the fit parameters are usually sensitive to the chosen fitting range,\cite{weysser2010structural} we have carefully fitted the data over a time domain
where the parameters are fairly robust and exhibit only a weak dependence on the fit boundaries. The $A(k)$ are very close to and only slightly 
smaller than the corresponding $f^c(k)$ in Fig.~\ref{fig:fc}, hence we do not show them here. For both the relaxation time $\tau_K(k)$ and the stretching 
exponent $\beta(k)$, we again see a convergence trend upon increasing the MF closure level $N$. Interestingly, the $\beta(k)$ exponent decreases as level $N$ increases. 
At large wavenumbers, $\beta(k)$ converges to $b$ for all levels; this property has been rigorously shown to hold in standard MCT,\cite{fuchs1994kohlrausch} 
and here we find that it is also correct in higher-order GMCT. 

Let us finally compare our GMCT predictions for the $\alpha$-relaxation of Percus-Yevick hard spheres to the simulation data of weakly polydisperse hard spheres.
Weysser \textit{et al.}\cite{weysser2010structural} demonstrated that the $\alpha$-relaxation process at $\varphi=0.585$ can be accurately 
fitted by the stretched-exponential Kohlrausch function. However, 
their empirically determined value of $\beta(k)\rightarrow0.5$ at large wavenumbers ($kd=20$) was found to be significantly lower than the standard MCT 
prediction of $b=0.61$, and this overestimation could not be remedied by incorporating more particle species in the polydisperse MCT analysis.
Furthermore, there was a notable difference between the fitted $\beta(k)$ over all wavenumbers and 
the standard MCT predicted $\beta(k)$ (see Fig.~8 in Ref.\ \onlinecite{weysser2010structural}). This undoubtedly reveals that standard MCT 
overestimates the $\beta(k)$ as well as the exponent $b$. Remarkably, our results systematically lower the $\beta(k)$ and $b$ when using a higher MF closure 
level $N$. In fact, if we only consider wavenumbers up to $kd=20$, the $\beta(k)$ in Fig.~\ref{fig:Kohlrausch}(b) for our highest closure level is 
very close to the simulation data in Fig.~8 in Ref \cite{weysser2010structural}. Hence, we may conclude that higher-order GMCT can quantitatively improve the stretched
exponents $\beta(k)$. However, notice that the shape of $\tau_K(k)$ is still similar to that obtained from one-component standard MCT 
in Fig.~7 in Ref.\ \onlinecite{weysser2010structural}. Weysser \textit{et al.}\cite{weysser2010structural} found that the explicit inclusion of 
polydispersity effects via multi-component MCT can significantly improve the predicted $k$-dependence of the $\alpha$-relaxation times 
$\tau_K(k)$ for polydisperse hard spheres, especially at low wavenumbers. We expect the same trend to apply in our higher-order GMCT framework, and 
future work is planned to extend the current theory to multi-component GMCT.

Taken together, the results of this study indicate that the GMCT scaling laws in both the $\beta$- and $\alpha$-relaxation regimes are essentially the same as 
those predicted by standard MCT, except for the fact that the exponent parameters $a$, $b$, $\gamma$, $\lambda$ and $\beta(k)$ quantitatively change with the closure level $N$. 
Hence, we can conclude that our higher-order GMCT framework preserves some of the most remarkable successes of standard MCT, including a first-principles
account for the von Schweidler law, the time-wavenumber factorization property in the $\beta$-relaxation regime, a time-temperature superposition principle, 
and an entirely non-trivial connection between the early $\beta$-relaxation process, the late $\beta$-relaxation process, and the final $\alpha$-relaxation process.
Furthermore, while our 
current analysis for monodisperse Percus-Yevick hard spheres precludes a stringent comparison with experiment or simulation, 
the fit parameters obtained from higher-order GMCT are found to be in good agreement with empirical studies of weakly polydisperse hard-sphere suspensions. 
Notably, the predicted power-law exponent $\gamma$ for the structural relaxation time, the von Schweidler exponent $b$, and the stretched exponents $\beta(k)$ appear 
to improve as more levels are included in the GMCT hierarchy, 
offering hope for an ultimately fully first-principles-based and quantitatively accurate prediction of glassy dynamics.

\section{Conclusions}

In this work, we have presented a detailed numerical analysis of the glassy dynamics of Percus-Yevick hard spheres using first-principles-based generalized mode-coupling theory.
Using only the static structure factor $S(k)$ at a given packing fraction $\varphi$ as input, this framework then predicts the full microscopic relaxation dynamics at the corresponding
state point through a hierarchy of coupled integro-differential equations. 
We have considered two different types of approximations to close the GMCT hierarchy, namely self-consistent mean-field and exponential closures, 
and find that these constitute an upper and 
lower bound for the predicted time-dependent dynamics, respectively. 
Consistent with earlier GMCT studies,\cite{mayer2006cooperativity,janssen2015microscopic} we also find that both closure families uniformly converge as more levels are incorporated into the theory.

Our mean-field closure predictions show that the inclusion of more levels in the GMCT hierarchy 
leads to a systematic increase in the value of the critical packing fraction $\varphi^c$, thus remedying the general tendency of standard MCT 
to overestimate a system's glassiness. Indeed,
the higher-order GMCT framework manifestly introduces more ergodicity-restoring relaxation processes, allowing the dynamics to remain supercooled-liquid-like 
over a substantial domain in the activated regime. These results suggest that higher-order GMCT can provide a means to circumvent the artifacts of standard MCT's
uncontrolled factorization approximation in a controlled manner.  
The predicted critical non-ergodicity parameters $f^c(k)$ also convergently increase with the mean-field closure level $N$; 
it must be noted, however, that the values of $f^c(k)$ are rather sensitive to the numerical accuracy with which the critical point is determined. The overall increase of $f^c(k)$
with $N$ cannot merely be attributed to a more pronounced input-microstructure at a higher $\varphi^c$, but rather stems from a non-trivial interplay between the $\varphi^c$-dependent 
changes in $S(k)$ and the increased complexity of the higher-order GMCT equations.   

At any given value of the packing fraction, we find that increasing the mean-field closure level always yields faster time-dependent relaxation dynamics, again confirming
that higher-order GMCT provides a systematic means to introduce more ergodicity-restoring fluctuations. 
Interestingly, however, after rescaling the predicted GMCT dynamics with respect to the corresponding critical point $\varphi^c$, an increase in $N$ generally leads to 
relatively \textit{slower} dynamics, as well as to a higher plateau value of the two-point density correlation function $\phi_1(k,t)$. This effect is concomitant to
the increase of $f^c(k)$ with $N$ and is consequently rooted in the inherent complexity of $N$-dependent GMCT. For the predicted $\alpha$-relaxation times $\tau$, 
all mean-field closures considered in this work ($N\leq5$) conform to a power-law divergence that is qualitatively similar to the standard MCT prediction. However, the power-law exponent
$\gamma$ is found to increase with $N$ and approaches the empirical result for weakly polydisperse hard spheres more closely than standard MCT. 
For the exponential closure series ($N\leq7$) we find that the relaxation time grows more weakly with $\varphi$, and becomes more non-Arrhenius-like with increasing $N$.
It must be noted however that our exponential-closure calculations are still relatively far from convergence at high packing fractions, precluding a definite conclusion
on the final growth behavior.
 
The analytic and asymptotic scaling laws in the $\beta$- and $\alpha$-relaxation regimes--including non-trivial scalings for the onset to 
and decay away from the $\beta$-relaxation plateau, 
the characteric time scaling for $\beta$-relaxation, time-wavenumber factorization, a time-temperature (or time-density) superposition principle, and Kohlrausch stretching in
the $\alpha$-relaxation regime--, are all found to be similar to those of standard MCT at all mean-field closure levels considered. Importantly, however, the corresponding
critical exponent parameters $a$, $b$, $\gamma$, $\lambda$, as well as the critical amplitudes $h(k)$ and Kolhrausch exponents $\beta(k)$, all become explicitly $N$-dependent in higher-order GMCT. 
Furthermore, within any given mean-field closure level $N$, the parameters $a$, $b$, and $\gamma$ are connected via non-trivial relations that are 
fully preserved at all values of $N$. Hence, we conclude that the higher-order GMCT framework inherits some of the most celebrated successes of standard MCT, 
namely the detailed analytic prediction of universal--and generally far-from-trivial--scaling laws in both the $\beta$- and $\alpha$-relaxation regimes.

From all the results above, we can confidently conclude that first-principles-based microscopic GMCT is capable of entering into the activated glassy regime 
$\varphi^c_{\text{MCT}}<\varphi<\varphi^c_{\text{GMCT}}\le\varphi_\text{g}$, a regime that is usually deemed inaccessible to standard MCT.\cite{langer2014theories}
Importantly, the systematic inclusion of higher-order density correlations within GMCT does not merely amount to a shift of the critical point, rendering the GMCT predictions 
fundamentally distinct from a conventional rescaled MCT analysis. The fact that GMCT can quantitatively improve the $\gamma$ parameter for dense hard spheres
also makes the theory promising for new first-principles studies on the microscopic origins of fragility. In view of the equivalence between the glass transition and 
the colloidal glass transition in the hard-sphere limit,\cite{xu2009equivalence} the conclusions presented here may also be applicable to other systems with repulsive potentials.
It remains to be explored, however, whether the current GMCT framework can ultimately provide a unified picture for both fragile and strong materials; in particular, a test for
a strong glass-former such as silica would be vital to establish if a fundamental difference between strong and fragile vitrification can emerge from high-order GMCT. 
While our two choices of high-order closure approximations can manifestly account for both Arrhenius- and super-Arrhenius behavior, neither the mean-field nor exponential 
closure series are currently fully converged at high packing fractions. 
Moreover, even if enhanced computational power would allow us to approach the $N\rightarrow\infty$ 
limit more closely, let us recall that the current version of GMCT still contains certain approximations--most notably the neglect of off-diagonal dynamic 
density correlations.\cite{janssen2015microscopic} Future studies should clarify to what extent these remaining approximations influence the overall   
relaxation dynamics, and whether the theory can also adequately account for other glassy phenomena such as dynamic heterogeneity and Stokes-Einstein violation. 
 


%
%

%

\begin{acknowledgments}
It is a pleasure to thank J\"{u}rgen Horbach for many stimulating discussions on this work.
We acknowledge the Netherlands Organisation for Scientific Research (NWO) for financial support through a START-UP grant. 
\end{acknowledgments}

\bibliography{paper1}

\end{document}